\documentclass[12pt]{article}
\usepackage{amsmath,amsfonts,amssymb,latexsym,graphicx} 

\newcommand{\publication}{(preprint hep-ph/0505033, KA--TP--06--2005)}

\title{\vspace*{-2.5cm}{\Large\bf
Merging gauge coupling constants without\\ Grand Unification}}
\author{\\
F.R. Klinkhamer\,\thanks{Email address:
frans.klinkhamer@physik.uni-karlsruhe.de}\\
{\normalsize
Institute for Theoretical Physics, University of Karlsruhe (TH),}\\
{\normalsize 76128 Karlsruhe, Germany}\vspace*{0.25cm}\\
and \vspace*{0.25cm}\\
G.E. Volovik\,\thanks{Email address: volovik@boojum.hut.fi}\\
{\normalsize Low Temperature Laboratory, Helsinki University of Technology,}\\
{\normalsize P.O. Box 2200, FIN--02015 HUT, Finland;}\\
{\normalsize L.D. Landau Institute for Theoretical Physics,
Russian Academy of Sciences,}\\
{\normalsize Kosygina 2, 119334 Moscow, Russia}}

\date{\vspace*{0.25cm}{\normalsize
\publication
}}

\begin{document}
\onecolumn

\maketitle
\begin{abstract}
{\noindent
The merging of the running couplings constants of
the weak, strong, and electromagnetic
fields does not require
the unification of these gauge fields at high energy.
It can, in fact, be the property of a general
fermionic system in which gauge bosons are not fundamental. }
\end{abstract}

\def\g{\kappa}
\def\half{{\frac{1}{2}}}
\def\L{{\mathcal L}}
\def\S{{\mathcal S}}
\def\d{{\mathrm{d}}}
\def\x{{\mathbf x}}
\def\v{{\mathbf v}}
\def\im{{\rm i}}
\def\etal{{\emph{et al\/}}}
\def\det{{\mathrm{det}}}
\def\tr{{\mathrm{tr}}}
\def\ie{{\emph{i.e.}}}
\def\eg{{\emph{e.g.}}}
\def\bnabla{\mbox{\boldmath$\nabla$}}
\def\Box{\kern0.5pt{\lower0.1pt\vbox{\hrule height.5pt width 6.8pt
 \hbox{\vrule width.5pt height6pt \kern6pt \vrule width.3pt}
 \hrule height.3pt width 6.8pt} }\kern1.5pt}
\def\HRULE{{\bigskip\hrule\bigskip}}
\def\be{\begin{equation}}
\def\ee{\end{equation}}
\def\implies{\Rightarrow}

\newpage
\section{Introduction}
\label{sec:introduction}

There are several lessons to be learnt from the example of
relativistic quantum fields emerging in condensed matter.
One of them is that the physical cutoff can be different for bosons and
fermions if the fermions are more fundamental than the bosons.
This occurs in superfluid $^3$He--A, where bosons are the collective modes
of the fermionic quantum vacuum and are composite objects made of
fermionic degrees of freedom \cite{VolovikBook}. The naive counting of
fermionic and bosonic contributions to the vacuum polarization
suggests that the anti-screening effect of charged bosons must dominate
over the screening effect of the fermionic vacuum and that, therefore,
the effective $SU(2)$ gauge field emerging in
$^3$He--A must experience asymptotic freedom \cite{asymptoticfreedom}.

However, this is not what happens in superfluid $^3$He--A. Instead, the
$SU(2)$ coupling constant shows the same zero-charge effect as the
Abelian $U(1)$ field.
\footnote{The term ``zero-charge effect''
refers to the infrared behavior, whereas
``asymptotic freedom'' refers to the ultraviolet behavior.}
The reason is the difference in cutoff scales for
bosons and fermions. As a result, the contribution of the fermions to the
logarithmically running coupling constant prevails, in spite of the
larger boson content. Actually, the hierarchy of cutoff scales in
$^3$He--A is such that the asymptotic-freedom contribution from the
gauge bosons just does not develop and the only contribution to the
vacuum polarization comes from the fermions.

Another important lesson from condensed-matter physics is that the
bare coupling constant is absent for emergent gauge fields
of a fermionic quantum vacuum.
The reason is simply that such gauge bosons cannot exist as free fields,
that is, without having fermions around to make the quantum vacuum. This
implies, in particular, that the entire gauge coupling constant comes from
vacuum polarization.

Here, we assume that the Standard Model of elementary
particle physics also has different physical cutoff
scales: the compositeness scale $E_{\rm c}$ which provides the cutoff
for the gauge bosons and the much higher ultraviolet cutoff $E_{\rm UV}$
for the fermions. Assuming that all three coupling constants of the
Standard Model come exclusively from vacuum polarization, we will find
that the most natural choice for the compositeness scale $E_{\rm c}$ is
the Planck scale $E_{\rm Planck} \approx 10^{19}\,\mathrm{GeV}$
(or, possibly, a  scale lower by a factor of about $10^{4}$),
while the ultraviolet cutoff scale $E_{\rm UV}$ will turn out to be
much larger than the Planck scale.

This second cutoff may be associated with the energy scale where Lorentz
invariance is violated, $E_{\rm UV} \sim E_{\rm Lorentz}$.
It has been claimed \cite{LorentzNonViolation} that cosmic-ray
observations
imply $E_{\rm Lorentz} > 10^{21}\,\mathrm{GeV}$,
assuming the absence of very small numerical factors in the
dispersion relations.\footnote{An explicit calculation of photon
propagation in a static background of randomly positioned wormholes
has shown how, in principle, small numerical factors could appear
in the photon dispersion relation \cite{KlinkhamerRupp},
but this calculation does not apply to fermions.}
Probably, $E_{\rm Lorentz}$ is even
larger. This would mean that the Planck cutoff is highly Lorentz
invariant and that the underlying symmetry of the fundamental structure is
itself the Lorentz symmetry, which
then protects the Lorentz invariance of the
effective low-energy physics \cite{QGfinetuning}.

If $E_{\rm Lorentz}\gg E_{\rm Planck}$, the topological Fermi-point
scenario of emergent relativistic quantum fields may
be relevant \cite{VolovikBook}. Specifically,
the integration over fermions with energy $E \lesssim E_{\rm Planck}$
occurs in the fully
relativistic region, where fermions are still close to the Fermi points
and, therefore, have gauge invariance and general covariance. As a
result, the induced effective action for the gauge and gravity fields is
automatically invariant.

The small ratio of cutoff parameters, $E_{\rm Planck}^2/E_{\rm Lorentz}^2$,
protects the Lorentz invariance of the known physical
laws. This would be in accordance with Bjorken's suggestion
\cite{Bjorken} that a
highly accurate relativistic quantum field theory
can only emerge if there is a small expansion parameter in the theory.

The merging of gauge coupling constants at high energy is usually
associated with Grand Unification of weak, strong, and electroweak
interactions into a larger gauge group with a single coupling constant
\cite{UnificationModel,Unification}.
The two-scale scenario discussed in the present Letter demonstrates
that the merging of running couplings could occur without
unification, it could very well be the natural property of an
underlying fermionic vacuum.

\section{Running couplings from two energy scales}
\label{sec:Runningcouplings}

Let us assume that the gauge fields of the Standard Model are
not fundamental but induced,
so that the three running coupling constants $g_i$ of the gauge group
$U(1)\times SU(2)\times SU(3)$ only come from vacuum polarization.
In other words, the fine structure constants $\alpha_i \equiv g_i^2 /4\pi$,
for $i=1,2,3$,
are completely determined by logarithms and have vanishing bare coupling
constants,
$1/\alpha_i^{(0)}=0$.

 If gauge bosons are fermion composites, the ultraviolet cutoff scale for
the vacuum polarization caused by fermions must be larger than the one
caused by gauge bosons. Let $E_{\rm UV}$ be the cutoff for the fermions
and $E_{\rm c}\ll E_{\rm UV}$
the compositeness scale which provides the cutoff energy for the gauge
bosons. Then, for energies above the electroweak scale but below the
compositeness
scale, one has at one loop (cf. Refs.~\cite{Unification,Weinberg}):
\begin{subequations}\label{runningalphas}
\begin{eqnarray}
\alpha_1^{-1}(E) &=& \frac{5N_F}{9\pi}~ \ln\frac{E_{\rm UV}^2}{E^2}~,
\label{runningHyper}
\\[2mm]
\alpha_2^{-1}(E) &=& \frac{N_F}{3\pi}~ \ln\frac{E_{\rm UV}^2}{E^2}
-\frac{11}{6 \pi}~ \ln \frac{E_{\rm c}^2}{E^2}~,
\label{runningWeak}
\\[2mm]
\alpha_3^{-1}(E) &=& \frac{N_F}{3\pi}~ \ln\frac{E_{\rm UV}^2}{E^2}
-\frac{11}{4 \pi}~ \ln\frac{E_{\rm c}^2}{E^2}~,
\label{runningStrong}
\end{eqnarray}
\end{subequations}
for $M_Z \ll E \ll E_{\rm c} \ll E_{\rm UV}$
and natural units with $\hbar=c=1$.
Here, $N_F$ is the number of fermion families contributing to the
positive screening (zero-charge) vacuum
polarization, whereas the negative anti-screening (asymptotic-freedom)
contribution come from the non-Abelian gauge bosons.

At the compositeness scale $E_{\rm c}$, the weak and strong inverse
couplings, as well as the hypercharge inverse coupling with a
factor $3/5$, approach the same value,
\begin{equation}
\frac{3}{5}\:\alpha_1^{-1}(E_{\rm c}) =\alpha_2^{-1}(E_{\rm
c})=\alpha_3^{-1}(E_{\rm c})=
 \frac{N_F}{3\pi}~ \ln\frac{E_{\rm UV}^2}{E_{\rm c}^2}~.
\label{runningalphas-atEc}
\end{equation}
Above the compositeness scale, the behavior depends on the details of
the dynamics. If the gauge bosons break up for $E>E_{\rm c}$,
the story ends here, at least as far as the gauge bosons are concerned.
If, on the other hand, the gauge bosons survive
but for some reason do not contribute to the vacuum polarization,
the couplings run together as
\begin{equation}
\frac{3}{5}\:\alpha_1^{-1}(E) =\alpha_2^{-1}(E)=\alpha_3^{-1}(E)=
 \frac{N_F}{3\pi}~ \ln\frac{E_{\rm UV}^2}{E^2}~,
\label{runningalphas-aboveEc}
\end{equation}
for $ E_{\rm c} \ll E \ll E_{\rm UV}$.
As discussed in the Introduction, a similar situation occurs in superfluid
$^3$He--A, with only fermions contributing to the polarization of the
vacuum. In this liquid, the running coupling constant of the effective
$SU(2)$ field behaves in exactly the same way as the one of the Abelian
$U(1)$ field, that is, it experiences the same zero-charge effect.
Of course, as the couplings $\alpha_i$ from
Eq.~(\ref{runningalphas-aboveEc}) grow with energy,
higher-order
contributions need to be added to the logarithm shown (cf.
Ref.~\cite{Weinberg}).

Let us, first, estimate the compositeness scale $E_{\rm c}$.
This can be done in the same way as the standard
calculation of the unification scale (cf. Ref.~\cite{Weinberg}), \ie, only
using the bosonic contributions to the running couplings. One then
obtains for the compositeness energy scale the same value as usually
assumed to hold for Grand Unified Theories (GUTs).\footnote{The reason is
that the right-hand sides of Eqs.~(\ref{runningalphas}abc) can be written
solely in terms of
$\ln (E_{\rm c}^2/E^2)$ and $\widetilde{\alpha}^{-1} \equiv N_F/(3\pi)\,
\ln (E_{\rm UV}^2 / E_c^2)$, with $E_{\rm c}$ and $\widetilde{\alpha}$
taking the role of the unification energy $E_\mathrm{GUT}$
and coupling constant $\alpha_\mathrm{GUT}$.}

Cancelling out the fermionic contributions from the right-hand sides
of Eqs.~(\ref{runningalphas}abc),
one finds two equations at the electroweak scale $M_Z$,
\begin{subequations}
\begin{eqnarray}
\alpha_2^{-1}(M_Z) - \alpha_3^{-1}(M_Z) &=&
\frac{11}{12\, \pi}~ \ln\frac{E_{\rm c}^2}{M_Z^2}~,\\[2mm]
\label{StrongMinusWeak}
\frac{3}{5}\,\alpha_1^{-1}(M_Z) - \alpha_2^{-1}(M_Z) &=&
\frac{11}{6\, \pi}~ \ln\frac{E_{\rm c}^2}{M_Z^2}~.
\label{WeakMinusHyper}
\end{eqnarray}
\end{subequations}
Extracting the combination
$1/\alpha_{Q}\equiv 1/\alpha_{1}+ 1/\alpha_{2}$
from these equations,
one obtains Eq.~(21.5.16) of Ref.~\cite{Weinberg}, which expresses
the logarithm in terms of the strong coupling  constant
$\alpha_3$ and the fine structure constant $\alpha_{Q}$
at the electroweak scale,
\begin{equation}
 \ln\frac{E_{\rm c}^2}{M_Z^2} = \frac{2\,\pi}{11\,\alpha_{Q}(M_Z)}\,
\left(1-\frac{8}{3}\; \frac{\alpha_{Q}(M_Z)}{\alpha_{3}(M_Z)}\right)~.
\label{Logarithm}
\end{equation}
Taking the numerical values $\alpha_3(M_Z)\approx 0.120$ and
$\alpha_{Q}(M_Z)\approx1/128$ at
$E=M_Z \approx 91.2\;\mathrm{GeV}$  \cite{Weinberg}, this gives
the following estimate:
\begin{equation}
\ln (E_{\rm c}^2/ M_Z^2) \approx 60.4~.
\label{Ec}
\end{equation}
The compositeness scale $E_{\rm c}$ is about $10^{15}\,\mathrm{GeV}$,
which is relatively close to the Planck energy scale
$E_{\rm Planck} \equiv \sqrt{\hbar\,c^5/G} \approx
1.22 \times 10^{19}\,\mathrm{GeV}$.

\begin{figure*}[p]
\begin{center}
\includegraphics[width=7.5cm]{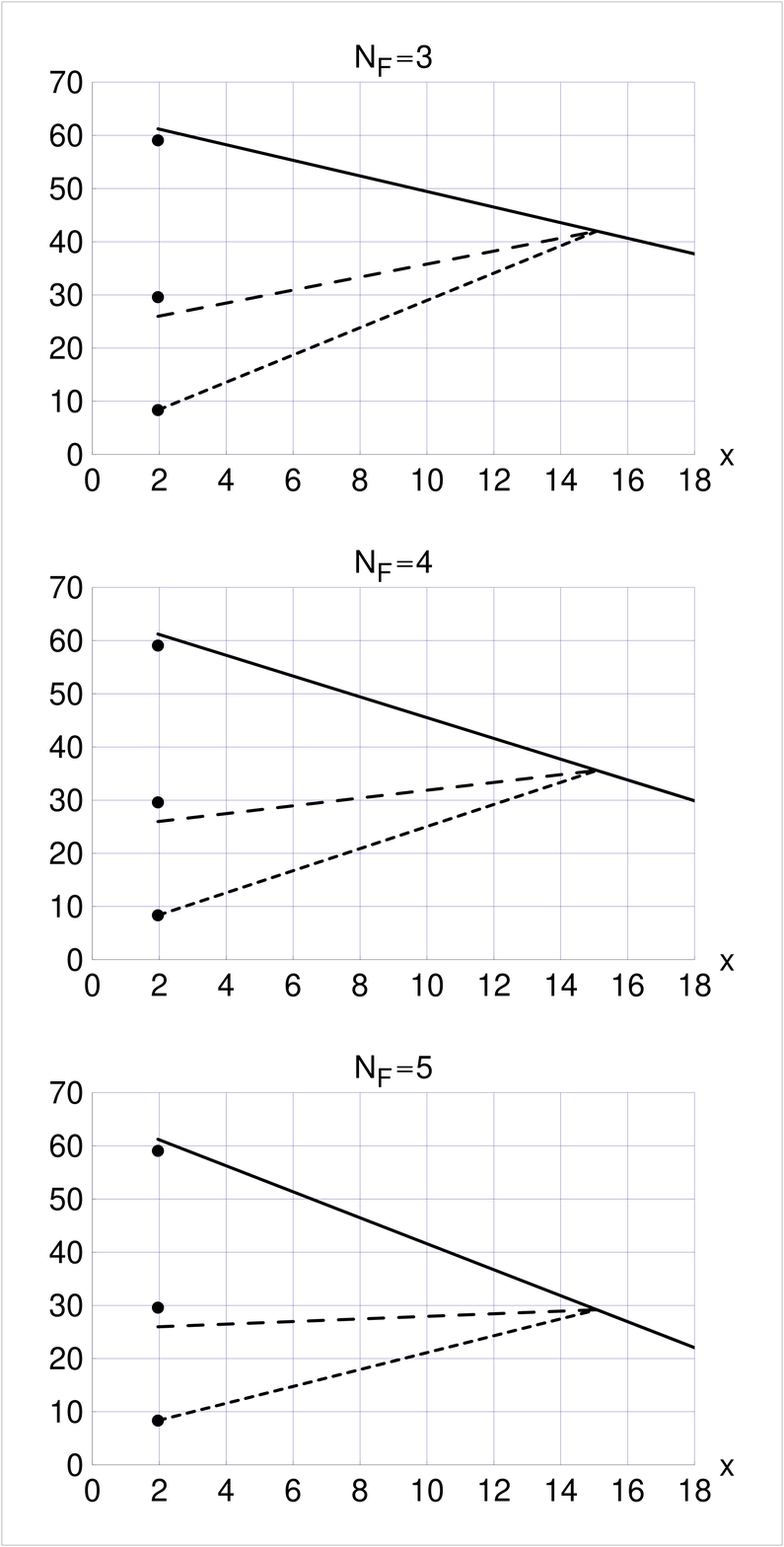}
\end{center}
\vspace*{-0.5cm}
\caption{Inverse couplings $(3/5) \times \alpha_1^{-1}$ [solid curve],
$\alpha_2^{-1}$ [long-dashed curve], and
$\alpha_3^{-1}$ [short-dashed curve], as a function of
$x \equiv \log_{10} (E/\mathrm{GeV})$
for different numbers $N_F$ of fermion families.
The coupling constants are given by
Eqs.~(\ref{runningalphas}), (\ref{runningalphas-aboveEc}),
(\ref{Ec}), (\ref{Euv}), and run together for $E>E_c$ [overlapping
curves].
At the compositeness scale $E \sim E_c \approx 10^{15}
\;\mathrm{GeV}$, there may be threshold effects which somewhat change the
values of the couplings towards lower energies (see text). The dots show
the experimental values at $E=M_Z \approx
91.2\;\mathrm{GeV}$.}
\label{mergingcouplingsFIG}
\end{figure*}

Let us now estimate the ultraviolet cutoff $E_{\rm UV}$ for the fermions.
From Eqs.~(\ref{runningHyper}) and (\ref{runningWeak}),    
the fine structure constant $\alpha_{Q}$ at the electroweak scale reads
\begin{equation}
\alpha_Q^{-1}(M_Z) =
\frac{8N_F}{9\pi}~ \ln\frac{E_{\rm UV}^2}{M_Z^2} -
\frac{11}{6 \pi}~ \ln\frac{E_{\rm c}^2}{M_Z^2}~.
\label{FineStructureEW2}
\end{equation}
Using Eq.~(\ref{Logarithm}) to eliminate the compositeness scale
$E_{\rm c}$,
one obtains
\begin{equation}
\ln\frac{E_{\rm UV}^2}{M_Z^2} =                   
\frac{3\,\pi}{2\,N_F\,\alpha_{Q}(M_Z)}\,
\left(1-\frac{2}{3}\; \frac{\alpha_{Q}(M_Z)}{\alpha_{3}(M_Z)}\right)~.
\label{LogarithmUV}
\end{equation}
With the numerical values mentioned above, this gives the following
estimate:
\begin{equation}
\ln (E_{\rm UV}^2/ M_Z^2)\approx 577/N_F~.
\label{Euv}
\end{equation}
For $N_F=3$, one has
$\ln (E_{\rm UV}^2/ M_Z^2)\approx 192$, so that $E_{\rm UV} \approx
10^{44} \,\mathrm{GeV} \gg E_{\rm Planck}$.
For $N_F=5$, the fermion scale is still
larger than the Planck energy by a factor $10^8$.
The corresponding running coupling constants are shown in
Fig.~\ref{mergingcouplingsFIG}.

We realize, of course, that the re\-nor\-ma\-li\-za\-tion-group equations
(\ref{runningalphas}abc), with the numerical values (\ref{Ec}) and
(\ref{Euv}) inserted, give a weak mixing
angle at $E=M_Z$ somewhat below the experimental value
(cf. Fig.~\ref{mergingcouplingsFIG}). Specifically, we
find $\sin^2 \theta_w \approx 0.203$ instead of the experimental
value $0.231$ \cite{Weinberg}.
Alternatively, adding appropriate bare coupling constants
$1/\alpha_i^{(0)}$ to the right-hand sides of
Eqs.~(\ref{runningalphas}abc)
in order to match the three experimental values at $E=M_Z$,
we do not find precisely merging coupling constants at high energy.

For a genuine Grand Unified Theory, the problem is serious and has
been addressed in different ways; see, e.g.,
Refs.~\cite{SusyGUTs,LEP-GUTs,Mohapatra-etal,Kawamura} and references
therein. But, for a dynamic scenario as ours, the precise definition of
the compositeness scale is rather uncertain. The scale can, in fact,
be slightly different for the various composite gauge bosons.
In other words, the three couplings of
our scenario need not merge exactly at one particular energy (for
example, two couplings could merge first and the third one later).

The simplest way to model these threshold effects
is to replace $E_{\rm c}$ in Eqs.~(\ref{runningWeak})
and (\ref{runningStrong}) by $E_{\rm c2}$ and $E_{\rm c3}$, respectively,
where $E_{\rm c2}$ and $E_{\rm c3}$ are assumed
to be not more than a few orders of magnitude away from the
geometric average  $E_{\rm c} \equiv \sqrt{E_{\rm c2}\,E_{\rm c3}}$.
The experimental values $\alpha^{\rm exp}_i(M_Z)$ then give
$\ln (E_{\rm c2}^2/ M_Z^2) \approx 50.5$,
$\ln (E_{\rm c3}^2/ M_Z^2) \approx 58.0$, and
$\ln (E_{\rm UV}^2/ M_Z^2)\approx  557 /N_F$.
This suggests that the  range for threshold effects in
$E_{\rm c}$ may be approximately $10^{13}-10^{15}\,\mathrm{GeV}$
(which is also clear from Fig.~\ref{mergingcouplingsFIG}
by making appropriate shifts of the curves).
Note, that, without grand-unified group, there is no danger
of having too rapid proton decay.

\section{Discussion}
\label{sec:Discussion}

Let us end with a few general remarks.  Trans-Planckian cutoff scales
have been considered before, for example the scale
$E_\mathrm{cutoff} \approx 10^{42}\,\mathrm{GeV}$ in
Ref.~\cite{Laperashvili} as corresponding to
an exotic (non-existing) case. The condensed-matter-like scenario
discussed in the present article suggests, however, that this possibility
must be taken seriously.

In this scenario, the merging of the running couplings of weak, strong,
and
hypercharge fields does not require a unification of these fields
at high energy, it may simply be the property of a
fermionic system in which gauge bosons are not fundamental.
The factor $3/5$ for $\alpha_1^{-1}$ in Eq.~(\ref{runningalphas-aboveEc})
may indicate an underlying continuous or discrete symmetry
between the fermion species.

The large separation between the Planckian (or near-Planckian)
compositeness scale $E_{\rm c}$ and the trans-Planckian scale
$E_{\rm UV}$ may be of importance to considerations
of the Standard Model symmetries as emergent phenomena. In
particular, this allows us to discuss gauge invariance as being an
emergent symmetry.

 In the topological Fermi-point
scenario of emergent relativistic fields \cite{VolovikBook},
the spectrum of fermionic
excitations near the Fermi point is linear:
fermions are chiral and obey the
relativistic Weyl equation.
In this scenario, bosonic excitations behave as effective gauge fields
interacting with Weyl fermions. This implies that gauge invariance
automatically emerges in the fermionic sector close to the Fermi point,
\ie, at $E\ll E_{\rm UV}$. The fermions induce gauge invariance for the
effective action of the composite vector fields. Since the
compositeness cutoff parameter $E_{\rm c}$ is well below $E_{\rm UV}$,
gauge invariance in the bosonic sector is valid throughout the
compositeness scale $E_{\rm c}$. Hence, the requirement suggested by
Veltman
\cite{Veltman} is fulfilled. Specifically, he concluded that, if gauge
bosons are composite, gauge invariance should remain valid both in the
infrared ($E\ll E_{\rm c}$) and ultraviolet ($E\gg E_{\rm c}$) regions.
The high accuracy of gauge invariance in the bosonic sector is then
determined by the small parameter $E_{\rm c}^2/E_{\rm UV}^2$, in
accordance with a suggestion of Bjorken
\cite{Bjorken}.

In the Fermi-point scenario, $E_{\rm UV}$ is the scale below which the
spectrum of fermionic excitations near the Fermi point is linear, \ie,
Lorentz invariance induced by the Fermi point is still obeyed. That is why
the Lorentz-violation scale must be approximately equal to or larger than
$E_{\rm UV}$. In turn, this implies that Lorentz invariance is
more
fundamental than the other physical laws and that we cannot expect to
observe its violation in the near future.

Applying the two-scale formalism to gravity, one finds that it
gives the wrong value for the gravitational coupling constant. If $E_{\rm
UV}$ is again used as the energy cutoff for the fermionic contributions to
Newton's constant, one obtains $G^{-1}\sim N_F \, E_{\rm UV}^2$ instead of
$G^{-1}\sim N_F \, E_{\rm  Planck}^2$.
It is not clear at the moment how to cure this problem.

We can only speculate that non-logarithmic (power-law) divergences must be
considered with great care. For example, the fourth order divergence, which
leads to a vacuum energy density (cosmological constant $\Lambda$) of
order $E_{\rm UV}^4$ or
$E_{\rm Planck}^4$, can be cancelled without fine-tuning, due to
the thermodynamic stability of the vacuum \cite{Annalen}. The same may
hold for the Higgs mass problem---controlling the
quadratically divergent
quantum corrections to the Higgs potential mass term (see, \eg,
Ref.~\cite{HiggsMassProblem}). This cutoff-dependent mass term is simply
absorbed by the vacuum energy density and is zero in the equilibrium
vacuum, again due to thermodynamic stability
\cite{Higgs}. For induced gravity, the cancellation of the vacuum energy
density is demonstrated by a calculation of $\Lambda$ on a
$(3+1)$-dimensional brane embedded in $\mathrm{AdS}_5$ space: the
induced cosmological constant on the brane vanishes without fine-tuning,
due to a cancellation of the contributions from
$(4+1)$-dimensional fermionic matter and gravity \cite{Andrianov}.

There may very well be a general principle from the
underlying physics, which protects against $E_{\rm UV}^n$
contributions to $G^{-1}$
with $n>0$. Let us mention, in this respect, another example of induced
Sakharov gravity in terms of constituent fields, namely
Ref.~\cite{FrolovFursaev}, which used such a principle and demonstrated the
advantage of two energy scales. In the scheme of
Ref.~\cite{FrolovFursaev}, the first (lowest) energy scale is
the mass scale $M^{\,\prime}$ of the constituent fields.
With $M^{\,\prime}\sim E_{\rm Planck}$,
this provides a natural cutoff which determines  Newton's
constant, $G^{-1}\sim (M^{\,\prime})^2 \sim E_{\rm Planck}^2$.
The much higher cutoff
$E_{\rm UV}^{\,\prime}$ drops out from the effective action due to
imposed cancellations between the constituent fields (see also Ref.
\cite{Veltman}, where cancellations of fermionic and bosonic effects
are required). This scheme only works if Lorentz invariance survives
beyond the Planck scale, again in agreement with the statement in
Ref.~\cite{Veltman} that the symmetry should remain valid throughout the
cutoff range. The higher cutoff $E_{\rm UV}^{\,\prime}$ of
Ref.~\cite{FrolovFursaev} must, therefore, be below the Lorentz-violation scale.

In conclusion, it is possible that the scenario of emergent physics,
in combination with a hierarchy of cutoff energy scales,
can replace the grand-unification scenario based on symmetry breaking.
This new scenario
(with parameters $N_F$ and $E_{\rm c}\ll E_{\rm UV}$)
naturally leads to the merging of gauge coupling constants, without
the need to introduce a simple gauge group
(and without having to worry about too rapid proton decay or
too many magnetic monopoles left over from the early universe).

Moreover, the hierarchy of cutoff energy scales may be related
to the well-known hierarchy problem of the Standard Model---the absence
of a natural explanation for having
$M_Z \ll E_\mathrm{GUT} \;\mathrm{or}\; E_{\rm Planck}$.
The $^3$He--A example mentioned in the Introduction,
where gauge invariance is not fundamental,
suggests that the mass of the weak vector bosons
may result not from spontaneous symmetry breaking but from terms
depending on the ultraviolet cutoff.
If we accept this viewpoint, the typical value of the weak
vector boson mass would be $M_Z\sim E^2_{\rm c}/E_{\rm UV} \ll E_{\rm c}$,
which would be a first step towards understanding the
Standard Model hierarchy problem
mentioned above (with $E_{\rm c}$ taking the place of
$E_\mathrm{GUT}$). From the numerical estimates given in
Eqs.~(\ref{Ec}) and (\ref{Euv})
and without further threshold effects at the cutoff energies,
the suggested hierarchy would seem to
prefer having more than $N_F=3$ fermion families.

\section*{Acknowledgements}   

The work of G.E.V. is supported in part
by the Russian Ministry of Education and Science,
through the Leading Scientific School grant $\#$2338.2003.2.
This work is also supported by the
European Science Foundation COSLAB Program.

\vspace{2\baselineskip}

\end{document}